\documentstyle[12pt]{article}
\def\href#1#2{{#2}}
\begin{document}
\begin{titlepage}
\begin{flushright}
hep-ph/9512438 \\
THEP-95-4 \\
December 1995
\end{flushright}
\vspace{0.2cm}
\begin{center}
\LARGE
$R_{b}$ and $R_{c}$ Crises \\
\vspace{1cm}
\normalsize
Frank D. (Tony) Smith, Jr. \\
\footnotesize
e-mail: gt0109e@prism.gatech.edu \\
and fsmith@pinet.aip.org  \\
P. O. Box for snail-mail: \\
P. O. Box 370, Cartersville, Georgia 30120 USA \\
\href{http://www.gatech.edu/tsmith/home.html}{WWW
URL http://www.gatech.edu/tsmith/home.html} \\
and\\
\href{http://galaxy.cau.edu/tsmith/TShome.html}{WWW
URL http://galaxy.cau.edu/tsmith/TShome.html} \\
\vspace{12pt}
School of Physics  \\
Georgia Institute of Technology \\
Atlanta, Georgia 30332 \\
\vspace{0.2cm}
\end{center}
\normalsize
\begin{abstract}
The $R_{b}$ and $R_{c}$ crises described by Kaoru Hagiwara
\newline
in \href{http://xxx.lanl.gov/abs/hep-ph/9512425}{hep-ph/9512425}
\cite{HAG}
can be resolved by the T-quark mass value of 130 GeV and
the $\alpha_{s}(M_{Z})$ value of 0.106 of the
$D_{4}-D_{5}-E_{6}$ model described in
\href{http://xxx.lanl.gov/abs/hep-ph/9501252}{hep-ph/9501252}
\cite{SMI7}
and
\href{http://xxx.lanl.gov/abs/quant-ph/9503009}{quant-ph/9503009}
\cite{SMI8}.
\end{abstract}
\vspace{0.2cm}
\normalsize
\footnoterule
\noindent
\footnotesize
\copyright 1995 Frank D. (Tony) Smith, Jr., Cartersville, Georgia USA
\end{titlepage}
\newpage
\setcounter{footnote}{0}
\setcounter{equation}{0}

\newpage

\section{Introduction.}

During 1995, precision electroweak data have confirmed
the predictions of the Standard Model, with no new physics,
with the possible exception of the two observables $R_{b}$ and
$R_{c}$.

\vspace{12pt}

In his recent review,
\href{http://xxx.lanl.gov/abs/hep-ph/9512425}{Kaoru Hagiwara} \cite{HAG}
has described the situation in detail, assuming the validity
of the CDF value of the Truth quark mass of about 175 GeV.

\vspace{12pt}

\href{http://xxx.lanl.gov/abs/hep-ph/9512425}{Hagiwara} \cite{HAG}
also notes that, although $\alpha_{s} \sim 0.12$ is favored from
electroweak data, some low-energy measurements favor lower values
$\alpha_{s} \sim 0.11$.  (For more extended discussion of
the $\alpha_{s}$ situation, see
\href{http://xxx.lanl.gov/abs/hep-ph/9511469}{Shifman} \cite{SHIF}.)

\vspace{12pt}

The purpose of this paper is to show that the data for $R_{b}$ and
$R_{c}$ are consistent with the Standard Model, as described by the
$D_{4}-D_{5}-E_{6}$ model \cite{SMI6,SMI7,SMI8},
without the need for new physics beyond the Standard Model
such as technicolor, extended technicolor, or
conventional supersymmetry.

\section{$R_{b}$ Crisis.}

\href{http://xxx.lanl.gov/abs/hep-ph/9512425}{Kaoru Hagiwara
in hep-ph/9512425} \cite{HAG} says that
precision electroweak data imply that $R_{b}$,
the partial $Z_{0}$ boson width ratio
of $b \overline{b}$ decay to total hadronic decay,
is about $3 \%$ larger than Standard Model predictions
with Truth quark mass in the range of the CDF value of about 175 GeV.

\vspace{12pt}

Donoghue, Golowich, and Holstein \cite{DGH} say (at p. 461)
\newline
"Interestingly, however, a more complete calculation reveals
\newline
a slight {\it{decrease}} to occur in the decay rate
$\Gamma_{Z_{0} \rightarrow b \overline{b}}$
\newline
as $m_{t}$ grows."

\vspace{12pt}

Since the $D_{4}-D_{5}-E_{6}$ model \cite{SMI6,SMI7,SMI8} predicts a
Truth quark Mass of about 130 GeV,
as opposed to the CDF value of about 175 GeV,
\newline
equation 1.17 on p. 436 of \cite{DGH}

\begin{equation}
\Delta \rho \simeq 0.006 \times \left( \frac{m_{t}}{140 GeV} \right)^{2}
\end{equation}

and equation 5.19 on p. 461 of \cite{DGH}

\begin{equation}
\left( \Delta \rho \right)_{nonuniv}^{b} = - \frac{4}{3} \Delta \rho -
\frac{\alpha}{4 \pi s_{w}^{2}}
\left( \frac{8}{3} + \frac{1}{6 c_{w}^{2}} \right)
ln \frac{m_{t}^{2}}{M_{W}^{2}}
\end{equation}

can be applied to the $D_{4}-D_{5}-E_{6}$ model \cite{SMI6,SMI7,SMI8}
to give an $R_{b}$ value about $2.5 \%$ larger than $R_{b}$ based
on the CDF value of the Truth quark mass.

\vspace{12pt}

This shows that the
$D_{4}-D_{5}-E_{6}$ model \cite{SMI6,SMI7,SMI8}
is consistent with the Standard Model value of $R_{b}$.

\section{$R_{c}$ Crisis.}

\href{http://xxx.lanl.gov/abs/hep-ph/9512425}{Hagiwara's Fig. 3}
\cite{HAG} shows that
if $R_{c}$, the partial $Z_{0}$ boson width ratio
of $c \overline{c}$ decay to total hadronic decay,
is fixed at the Standard Model value,
then the precision electroweak data imply that $\alpha_{s}(M_{Z})$,
the color force coupling constant at the energy of the $Z_{0}$ mass,
is about $0.104 \pm 0.08$.

\vspace{12pt}

Since the $D_{4}-D_{5}-E_{6}$ model \cite{SMI6,SMI7,SMI8}
value of $\alpha_{s}(M_{Z})$ is about 0.106,
the $D_{4}-D_{5}-E_{6}$ model \cite{SMI6,SMI7,SMI8}
is consistent with the Standard Model value of $R_{c}$.

\end{document}